\documentclass[12pt]{article}

\usepackage{amsmath}
\usepackage{amssymb}
\usepackage{amsfonts}
\usepackage{graphics}

\DeclareMathOperator{\sech}{sech}

\title{
Semi-infinite quantum wells in a position-dependent mass background}
\author{C. Quesne\thanks{e-mail: Christiane.Quesne@ulb.be}\\
{\small\sl Physique Nucl\'eaire Th\'eorique et Physique Math\'ematique,  Universit\'e Libre de Bruxelles,} \\ 
{\small\sl Campus de la Plaine CP229, Boulevard~du Triomphe, B-1050 Brussels, Belgium}}
\date{ }
\begin{document}
\baselineskip=22pt plus 1pt minus 1pt
\maketitle

\begin{abstract} 
By using a point canonical transformation starting from the constant-mass Schr\"odinger equation for the Morse potential, it is shown that a semi-infinite quantum well model with a non-rectangular profile associated with a position-dependent mass that becomes infinite for some negative value of the position, while going to a constant for a large positive value of the latter, can be easily derived. In addition, another type of semi-infinite quantum well associated with the same position-dependent mass is constructed and solved by starting from the Rosen-Morse II potential instead of the Morse one.
\end{abstract}


\noindent
MSC Nos.: 81Q05; 81Q80
%
%
\newpage

\section{Introduction}

Due to its many applications in several fields of physics, the Schr\"odinger equation in a position-dependent mass  (PDM) background has arisen much interest over the years \cite{bastard, weisbuch, serra, harrison, barranco, geller, arias, puente, ring, bonatsos, willatzen, chamel}. Finding exact solutions of such a Schr\"odinger equation turns out to be very useful for understanding some physical phenomena and testing some approximation methods. Several techniques are available for generating PDM and potential pairs leading to such exact solutions (see, e.g., \cite{cq04, 
bagchi05, cq06} and references quoted therein). Among them, one may quote the point canonical transformation (PCT) applied to an exactly-solvable constant-mass Schr\"odinger equation \cite{bagchi04}, which has proved very useful to achieve this goal (see, e.g., \cite{cq09, cq21, cq22}).\par
%
%
In a series of papers, Jafarov, Nagiyev, and their collaborators studied several variants of harmonic oscillator models in a PDM background \cite{jafarov20a, jafarov20b, jafarov21a, jafarov21b, jafarov21c, jafarov21d, nagiyev22a, jafarov22, nagiyev22b}. They directly solved the corresponding Schr\"odinger equations and proved that their solutions go over to the well-known ones of the constant-mass harmonic oscillator in some appropriate parameter limit. In a recent work \cite{jafarov21e}, they considered a harmonic oscillator potential well with a PDM that has the property of becoming infinite for some negative value $-a$ of the position while going to a constant for a large positive value of the latter. This resulted in a semi-infinite quantum well, whose spectrum was determined by a direct approach.\par
%
%
The purpose of the present paper is twofold: first to demonstrate that the PCT method provides a simple derivation of some of the results of Ref.~\cite{jafarov21e}, and second to show that the same PDM may provide other exactly-solvable semi-infinite quantum wells.\par
%
%
This paper is organized as follows. In Sect.~2, the model of Ref.~\cite{jafarov21e} is reviewed and shown to be derivable from the known results for the Morse potential with a constant mass \cite{morse, cooper}. In Sect.~3, another type of semi-infinite quantum well is constructed and solved. Finally, Sect.~4 contains the conclusion.\par
%
%
\section{The Jafarov and Nagiyev semi-infinite quantum well and its derivation by the PCT technique}

In Ref.~\cite{jafarov21e}, Jafarov and Nagiyev considered the PDM Schr\"odinger equation
\begin{equation}
  \left(- \frac{d}{dx} \frac{1}{M(x)} \frac{d}{dx} + V_{\rm eff}(x)\right) \psi(x) = E \psi(x),  \label{eq:SE-PDM}
\end{equation}
where $M(x)$ and $V_{\rm eff}(x)$ are defined by
\begin{equation}
  M(x) = \left(1+\frac{x}{a}\right)^{-2}, \quad V_{\rm eff}(x) = \frac{1}{4} M(x) \omega^2 x^2 = 
  \frac{1}{4} \frac{a^2\omega^2 x^2}{(x+a)^2}, \quad -a < x < +\infty,  \label{eq:M}
\end{equation}
with $a>0$.\footnote{Note that we have adopted here units wherein $\hbar = 2m_0 = 1$ in the original paper.} Since the mass and the potential go to $+\infty$ for $x\to -a$, the definition of the latter may be completed by setting $V_{\rm eff}(x) = + \infty$ for $x\le-a$. On the other hand, for $x\to + \infty$, it is clear that $V_{\rm eff}(x) \to \frac{1}{4}\omega^2 a^2$. The potential is therefore characterized by an infinite wall on the left and a finite wall on the right. Hence, it is a semi-infinite quantum well with a non-rectangular profile.\par
%
%
At this stage, one may observe that the BenDaniel-Duke form \cite{bendaniel} of the kinetic energy operator adopted in (\ref{eq:SE-PDM}) is only a special case of the von Roos general two-parameter form \cite{vonroos}. Other orderings of the mass and the differential operator, such as the Zhu-Kroemer \cite{zhu} or the Mustafa-Mazharimousavi \cite{mustafa07, mustafa19} ones, which pass the de Souza Dutra and Almeida test \cite{dutra}, might have been chosen, but, as shown elsewhere \cite{cq22}, in general they do not change the results much.\par
%
%
In Ref.~\cite{jafarov21e}, the bound-state solutions of eq.~(\ref{eq:SE-PDM}) were shown to correspond to the energies
\begin{equation}
  E_n = \omega\left(n + \frac{1}{2}\right) - \frac{1}{a^2}n(n+1), \qquad n=0, 1, \ldots, N, \qquad N < \frac{1}
  {2}(\omega a^2 - 1).  \label{eq:E}
\end{equation}
The corresponding wavefunctions were obtained in the form
\begin{equation}
  \psi_n(x) \propto (x+a)^{-\frac{1}{2}\omega a^2 + n} \exp\left(- \frac{\omega a^3}{2(x+a)}\right)
  L_n^{(\omega a^2 - 2n -1)}\left(\frac{\omega a^3}{x+a}\right),
\end{equation}
expressed in terms of Laguerre polynomials $L_n^{(\alpha)}(z)$ and vanishing at $x=-a$ and $x\to+\infty$. This form was then transformed into
\begin{equation}
  \psi_n(x) \propto (x+a)^{-\frac{1}{2}\omega a^2} \exp\left(-\frac{\omega a^3}{2(x+a)}\right) 
  y_n\left(2\frac{x+a}{\omega a^3}; -\omega a^2\right),
\end{equation}
where $y_n(z;\alpha)$ denotes a Bessel polynomial. It is this latter form whose normalization coefficient was calculated and that was shown to reduce to the well-known wavefunctions of the constant-mass harmonic oscillator in the $a\to+\infty$ limit. For $E > \frac{1}{4}\omega^2 a^2$, eq.~(\ref{eq:SE-PDM}) also has a continuous spectrum, whose wavefunctions were briefly discussed in \cite{jafarov21e}.\par
%
%
These results may be alternatively obtained by using the PCT technique applied to a constant-mass Schr\"odinger equation \cite{bagchi04}
\begin{equation}
  \left(- \frac{d^2}{du^2} + U(u)\right) \phi(u) = \epsilon \phi(u) \label{eq:SE}
\end{equation}
for an appropriately chosen potential $U(u)$, defined on a finite or infinite interval. Such a method uses a change of variable
\begin{equation}
  u(x) = \bar{a} v(x) + \bar{b}, \qquad v(x) = \int^x \sqrt{M(x')}\, dx',  \label{eq:u-v}
\end{equation}
and a change of function
\begin{equation}
  \phi(u(x)) \propto [M(x)]^{-1/4} \psi(x).  \label{eq:phi-psi}
\end{equation}
Here, $\bar{a}$ and $\bar{b}$ are assumed to be two real parameters. The potential $V_{\rm eff}(x)$, defined on a possibly different interval, and the bound-state energies $E_n$ of the PDM Schr\"odinger equation are given in terms of the potential $U(u)$ and the bound-state energies $\epsilon_n$ of the constant-mass one by
\begin{equation}
  V_{\rm eff}(x) = \bar{a}^2 U(u(x)) + \frac{M^{\prime\prime}}{4M^2} - \frac{7M^{\prime 2}}{16M^3} + \bar{c},
  \label{eq:V-U}
\end{equation}
and
\begin{equation}
  E_n = \bar{a}^2 \epsilon_n + \bar{c},  \label{eq:E-epsilon}
\end{equation}
where a prime denotes derivation with respect to $x$ and $\bar{c}$ is some additional real constant. From (\ref{eq:phi-psi}), the corresponding bound-state wavefunctions are given by
\begin{equation}
  \psi_n(x) = \lambda [M(x)]^{1/4} \phi_n(u(x)) \label{eq:psi-phi}
\end{equation}
provided they are normalizable on the definition interval of $x$. Here $\lambda$ is some constant which may arise from the change of normalization when going from $u$ to $x$.\par
%
%
{}For the mass chosen in (\ref{eq:M}), one finds that the mass-dependent term in (\ref{eq:V-U}) is given by
\begin{equation}
  \frac{M^{\prime\prime}}{4M^2} - \frac{7M^{\prime2}}{16M^3} = - \frac{1}{4a^2}. \label{eq:M-dep}
\end{equation}
and that the change of variable (\ref{eq:u-v}) writes
\begin{equation}
  u(x) = \bar{a} a \log(x+a) + \bar{b}.  \label{eq:u-v-bis}
\end{equation}
On assuming
\begin{equation}
  \bar{a} = \frac{1}{a}, \qquad \bar{b} = 0,  \label{eq:parameter}
\end{equation}
the latter reduces to
\begin{equation}
  u(x) = \log(x+a) \qquad \text{or} \qquad e^{u(x)} = x+a.  \label{eq:u(x)}
\end{equation}
\par
%
%
{}From (\ref{eq:u(x)}), it looks appropriate to assume for the potential $U(u)$ in the starting constant-mass Schr\"odinger equation (\ref{eq:SE}) some function of the real exponential $e^u$. Let us consider the Morse potential \cite{morse}, which may be written as \cite{cooper}
\begin{equation}
  U(u) = B^2 e^{-2u} - B(2A+1) e^{-u}, \qquad -\infty < u < + \infty,  \label{eq:Morse}
\end{equation}
where $A$ and $B$ are two positive constants. Its bound-state eigenvalues are given by
\begin{equation}
  \epsilon_n = - (A-n)^2, \qquad n = 0, 1, \ldots, n_{\rm max}, \qquad n_{\rm max} < A,  \label{eq:E-Morse}
\end{equation}
and the corresponding eigenfunctions are
\begin{equation}
  \phi_n(u) = {\cal N}_n \exp[-(A-n)u - Be^{-u}] L_n^{(2A-2n)}(2Be^{-u}),
\end{equation}
with a normalization constant
\begin{equation}
  {\cal N}_n = (2B)^{A-n} \left(\frac{n! (2A-2n)}{\Gamma(2A+1-n)}\right)^{1/2}.  \label{eq:norm-Morse}
\end{equation}
\par
%
%
On applying eqs.~(\ref{eq:V-U}), (\ref{eq:M-dep}), (\ref{eq:parameter}), and (\ref{eq:u(x)}), potential (\ref{eq:Morse}) gets transformed into
\begin{equation}
  V_{\rm eff}(x) = \frac{1}{a^2} \left(\frac{B^2}{(x+a)^2} - \frac{B(2A+1)}{x+a}\right) - \frac{1}{4a^2} +
  \bar{c},
\end{equation}
which reduces to the potential $V_{\rm eff}(x)$ given in (\ref{eq:M}), provided we choose
\begin{equation}
  A = \frac{1}{2}(\omega a^2 - 1), \qquad B = \frac{1}{2} \omega a^3, \qquad \bar{c} = \frac{1}{4}\omega^2
  a^2 + \frac{1}{4a^2}.  \label{eq:AB}
\end{equation}
From (\ref{eq:E-epsilon}) and (\ref{eq:E-Morse}), its bound-state energies turn out to be given by (\ref{eq:E}) with $N = n_{\rm max}$, as it should be. From (\ref{eq:psi-phi}), the corresponding wavefunctions are obtained in the form
\begin{equation}
  \psi_n(x) = N_n (x+a)^{-\frac{1}{2}\omega a^2 + n} \exp\left(- \frac{\omega a^3}{2(x+a)}\right) L_n
  ^{(\omega a^2-2n-1)}\left(\frac{\omega a^3}{x+a}\right) \label{eq:wf}
\end{equation}
with
\begin{equation}
  N_n = \lambda \sqrt{a}\, {\cal N}_n
\end{equation}
and
\begin{equation}
  {\cal N}_n = (\omega a^3)^{\frac{1}{2}(\omega a^2-1)-n} \left(\frac{n! (\omega a^2 - 1 - 2n)}{\Gamma
  (\omega a^2 - n)}\right)^{1/2},  \label{eq:norm-Morse-bis}
\end{equation}
coming from (\ref{eq:norm-Morse}) and (\ref{eq:AB}). The remaining undetermined constant $\lambda$ can be easily found by making the change of variable $e^u = x+a$ in the normalization integral for $\psi_n(x)$ and using the known normalization coefficient ${\cal N}_n$ of $\psi_n(u(x))$. The result is $\lambda = 1/\sqrt{a}$, so that   $N_n$ of eq.~(\ref{eq:wf}) simply reduces to ${\cal N}_n$ given in (\ref{eq:norm-Morse-bis}).\par
%
%
We have therefore proved that the bound states of the semi-infinite quantum well with a non-rectangular profile of \cite{jafarov21e} can be easily derived from those of the constant-mass Morse potential. Its continuous spectrum eigenfunctions might also be obtained from those of the latter.\par
%
%
\section{Construction of another semi-infinite quantum well by the PCT technique}

In the present section, we will assume the same mass $M(x)$ as in (\ref{eq:M}), which implies that eqs.~(\ref{eq:M-dep}) and (\ref{eq:u-v-bis}) remain valid. With the same choice as in (\ref{eq:parameter}), we then get eq.~(\ref{eq:u(x)}).\par
%
%
Another potential for which the constant-mass Schr\"odinger equation is exactly solvable and that can be written in terms of the exponential $e^u$ is the Rosen-Morse II potential \cite{rosen}
\begin{equation}
  U(u) = - A(A+1) \sech^2 u + 2B \tanh u, \qquad -\infty < u < +\infty, \qquad B<A^2,  \label{eq:RM}
\end{equation}
whose bound-state energies and wavefunctions are given by \cite{cooper}
\begin{equation}
  \epsilon_n = - (A-n)^2 - \frac{B^2}{(A-n)^2}, \qquad n=0, 1, \ldots, n_{\rm max}, \qquad n_{\rm max}< A
  - \sqrt{|B|},
\end{equation}
and
\begin{align}
  \phi_n(x) &= {\cal N}_n (1-\tanh u)^{\left(A-n + \frac{B}{A-n}\right)/2} (1+ \tanh u)^{\left(A-n - \frac{B}{A-n}
       \right)/2} \nonumber \\
  & \quad \times P_n^{\left(A-n+\frac{B}{A-n}, A-n - \frac{B}{A-n}\right)}(\tanh u),
\end{align}
where $P_n^{(\alpha,\beta)}(z)$ denotes a Jacobi polynomial and ${\cal N}_n$ is a normalization coefficient given by \cite{levai}
\begin{equation}
  {\cal N}_n = 2^{n-A} \left(\frac{n! \Gamma(2A-n+1) [(A-n)^2 - B^2/(A-n)^2]}{(A-n) \Gamma(A+1+B/(A-n))
  \Gamma(A+1-B/(A-n))}\right)^{1/2}.  \label{eq:norm-RM}
\end{equation}
\par
%
%
In the special case where $B=0$, eqs.~(\ref{eq:RM})--(\ref{eq:norm-RM}) reduce to
\begin{equation}
  U(u) = - A(A+1) \sech^2 u,\qquad -\infty < u < + \infty,  \label{eq:RM-bis}
\end{equation} 
\begin{equation}
  \epsilon_n = - (A-n)^2, \qquad n=0, 1, \ldots, n_{\rm max}, \qquad n_{\rm max} < A, 
\end{equation}
\begin{equation}
  \phi_n(u) = \bar{{\cal N}}_n (\sech u)^{A-n} C_n^{\left(A-n+\frac{1}{2}\right)}(\tanh u),
\end{equation}
\begin{equation}
  \bar{{\cal N}}_n = \frac{\Gamma(2A-2n+1)}{2^{A-n}\Gamma(A-n+1)} \left(\frac{(A-n) n!}{\Gamma(2A-n+1)}
  \right)^{1/2},
\end{equation}
where the Jacobi polynomials with equal parameters have been expressed in terms of Gegenbauer ones using eq.~(22.5.20) of Ref.~\cite{abramowitz} and the new normalization factor $\bar{\cal N}_n$ combines the factor occurring in the latter equation with the normalization factor (\ref{eq:norm-RM}).\par
%
%
On applying eqs.~(\ref{eq:V-U}), (\ref{eq:M-dep}), (\ref{eq:parameter}), and (\ref{eq:u(x)}) to potential (\ref{eq:RM-bis}), we obtain
\begin{equation}
  V_{\rm eff}(x) = V_0 \left(\frac{1}{[(x+a)^2+1]^2} - \frac{1}{(x+a)^2+1}\right), \qquad -a < x < +\infty,
  \label{eq:potential-1}
\end{equation}
where we have set
\begin{equation}
  \bar{c} = \frac{1}{4a^2}, \qquad V_0 = \frac{4A(A+1)}{a^2} > 0.  \label{eq:V0}
\end{equation}
We note that
\begin{equation}
  \lim_{x\to -a} V_{\rm eff}(x) = \lim_{x\to+\infty} V_{\rm eff}(x) = 0
\end{equation}
and we complete the definition of $V_{\rm eff}(x)$ by setting
\begin{equation}
  V_{\rm eff}(x) = + \infty \qquad \text{for} \qquad x<-a,  \label{eq:potential-2}
\end{equation}
which agrees with the behaviour of the mass $M(x)$. On the interval $(-a, + \infty)$, the potential is negative and has a minimum given by
\begin{equation}
  V_{\rm eff}(x_{\rm min}) = - \frac{V_0}{4} \qquad \text{for} \qquad x_{\rm min} = -a+1.
\end{equation}
It is therefore another semi-infinite quantum well associated with the mass defined in (\ref{eq:M}). An example of such a potential is displayed in Fig.~1.\par
%
%
\begin{figure}
\begin{center}
\includegraphics{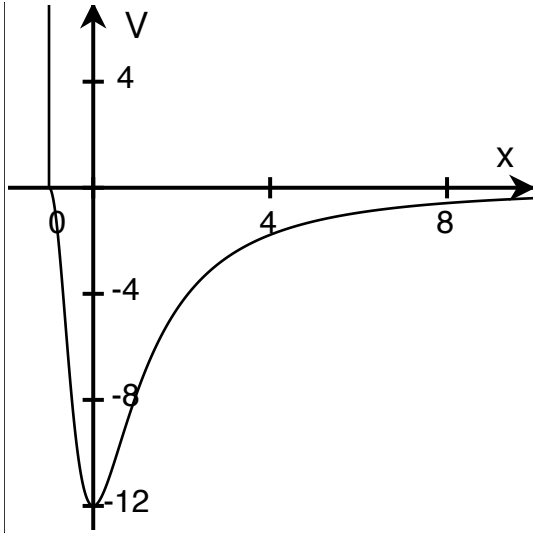}
\caption{Plot of the semi-infinite quantum well defined in (\ref{eq:potential-1}) and (\ref{eq:potential-2}) in terms of $x$ for $a=1$ and $V_0=48$.}
\end{center}
\end{figure}
\par
%
%
{}From eq.~(\ref{eq:V0}), it turns out that $A$ may be written as
\begin{equation}
  A = \nu - \frac{1}{2}, \qquad \nu \equiv \frac{1}{2} \sqrt{1 + a^2 V_0},
\end{equation}
so that on using eqs.~(\ref{eq:E-epsilon}), (\ref{eq:parameter}), and (\ref{eq:V0}), we obtain that the bound-state spectrum of $V_{\rm eff}(x)$ is given by
\begin{equation}
  E_n = - \frac{1}{a^2} (\nu-n) (\nu-n-1), \qquad n=0, 1, \ldots, n_{\rm max},
\end{equation}
while the corresponding wavefunctions can be expressed as 
\begin{equation}
  \psi_n(x) = N_n \frac{(x+a)^{\nu-n-1}}{[(x+a)^2+1]^{\nu-n-\frac{1}{2}}} C_n^{(\nu-n)}\left(\frac{(x+a)^2-1}
  {(x+a)^2+1}\right), \qquad -a < x < +\infty.
\end{equation}
Provided $n_{\rm max}$ is restricted to values such that $n_{\rm max}<\nu-1$, these wavefunctions vanish for $x\to -a$ and $x\to\infty$, as it should be. Their normalization coefficient is given by
\begin{equation}
  N_n = \lambda \sqrt{a} 2^{\nu-\frac{1}{2}-n} \bar{\cal N}_n,
\end{equation}
where the remaining undetermined constant $\lambda$ can be obtained by making the change of variable $e^u = x+a$ in the normalization integral of $\psi_n(x)$ on $(-a,+\infty)$. The result reads $\lambda = 1/\sqrt{a}$, so that
\begin{equation}
  N_n = \frac{\Gamma(2\nu-2n)}{\Gamma(\nu-n-\frac{1}{2})} \left(\frac{n!}{(\nu-n-\frac{1}{2}) \Gamma(2\nu-n)
  }\right)^{1/2}.
\end{equation}
\par
%
%
As an example, we show in Fig.~2 the three bound-state wavefunctions of the semi-infinite quantum well plotted in Fig.~1,
\begin{align}
  \psi_0(x) &= 2\sqrt{15} \frac{(x+1)^{5/2}}{[(x+1)^2+1]^3}, \label{eq:psi0}\\
  \psi_1(x) &= 2\sqrt{15} \frac{(x+1)^{3/2}}{[(x+1)^2+1]^3} [(x+1)^2-1], \label{eq:psi1}\\
  \psi_2(x) &= 2\sqrt{3} \frac{(x+1)^{1/2}}{[(x+1)^2+1]^3} [(x+1)^4 - 3(x+1)^2 + 1],  \label{eq:psi2}
\end{align}
corresponding to the energies $E_0 = - \frac{35}{4}$, $E_1 = - \frac{15}{4}$, and $E_2 = - \frac{3}{4}$, respectively.\par
%
%
\begin{figure}
\begin{center}
\includegraphics{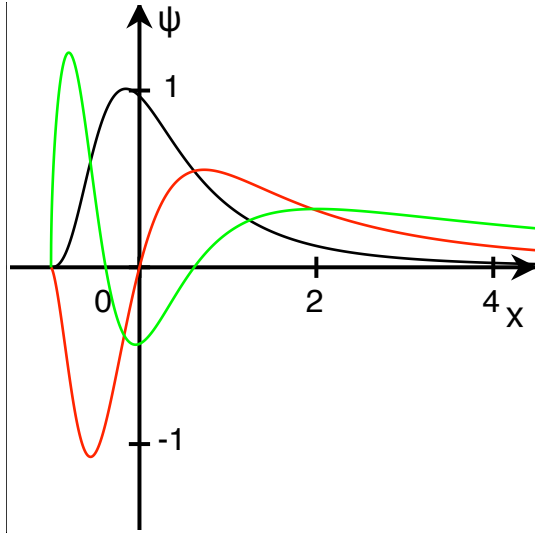}
\caption{Plot of $\psi_n(x)$, defined in (\ref{eq:psi0}), (\ref{eq:psi1}), and (\ref{eq:psi2}), in terms of $x$ for $n=0$ (black line), $n=1$ (red line), and $n=2$ (green line).}
\end{center}
\end{figure}
\par
%
%
\section{Conclusion}

In the present paper, we have first shown that the PCT method applied to the constant-mass Schr\"odinger equation  for the Morse potential allows us to easily retrieve some of the results of Ref.~\cite{jafarov21e}. In addition, we have proved that normalized bound-state wavefunctions can be directly expressed in terms of Laguerre polynomials, without having to transform them in terms of Bessel polynomials.\par
%
%
In a second step, we have derived another example of semi-infinite quantum well by starting from the Rosen-Morse II potential with $B=0$ and applying the same technique.\par
%
%
Considering some other exactly solvable potentials with constant mass as starting potential with the PDM defined in (\ref{eq:M}) may be an interesting topic for future investigation.\par
%
%
\section*{Acknowledgements}

This work was supported by the Fonds de la Recherche Scientifique-FNRS under Grant No.~4.45.10.08.\par
%
%
\section*{Data availability statement}

All data supporting the findings of this study are included in the article.\par
%
%
\section*{Authors declarations}

\section*{Conflicts of interest}

The author declares no conflict of interest.\par
%
%

\end{document}